\documentstyle[epsfig,12pt]{article}

\newskip\humongous \humongous=0pt plus 1000pt minus 1000pt

\newif\ifdtup

\def\Im{\mathop{\rm Im}}
\def\Re{\mathop{\rm Re}}

\let\vev\VEV

\def\ltap{\raisebox{-.4ex}{\rlap{$\sim$}} \raisebox{.4ex}{$<$}}

\newcommand{\MeV}{\rm\,MeV}

\def\pl#1#2#3{{\it Phys. Lett. }{\bf B#1~}#3~(19#2)}
\def\np#1#2#3{{\it Nucl. Phys. }{\bf B#1~}#3~(19#2)}

\makeatletter
\newcounter{alphaequation}[equation]
 \def\thealphaequation{\theequation\hbox to
0.6em{\hfil\alph{alphaequation}\hfil}}
\def\eqnsystem#1{
\def\@eqnnum{{\rm (\thealphaequation)}}
\def\@@eqncr{\let\@tempa\relax
\ifcase\@eqcnt \def\@tempa{& & &}
\or \def\@tempa{& &}\or \def\@tempa{&}\fi\@tempa
\if@eqnsw\@eqnnum\refstepcounter{alphaequation}\fi
\global\@eqnswtrue\global\@eqcnt=0\cr}
\refstepcounter{equation}
\let\@currentlabel\theequation
\def\@tempb{#1}
\ifx\@tempb\empty\else\label{#1}\fi
\refstepcounter{alphaequation}
\let\@currentlabel\thealphaequation
\global\@eqnswtrue\global\@eqcnt=0
\tabskip\@centering\let\\=\@eqncr
$$\halign to \displaywidth\bgroup
  \@eqnsel\hskip\@centering
  $\displaystyle\tabskip\z@{##}$&\global\@eqcnt\@ne
  \hskip2\arraycolsep\hfil${##}$\hfil&
  \global\@eqcnt\tw@\hskip2\arraycolsep
  $\displaystyle\tabskip\z@{##}$\hfil
  \tabskip\@centering&\llap{##}\tabskip\z@\cr}

\def\endeqnsystem{\@@eqncr\egroup$$\global\@ignoretrue}
\makeatother

\begin{document}
\begin{titlepage}
\begin{center}
October 1996 \hfill IFUP--TH 61/96 \\
~{} \hfill LBL--39488 \\
~{} \hfill OHSTPY--HEP--T--96--033 \\
~{} \hfill UCB--PTH--96/45  \\

\vskip .25in

{\large \bf Unified Theories With U(2) Flavor Symmetry}\footnote{This
work was supported in part by the Director, Office of
Energy Research, Office of High Energy and Nuclear Physics, Division of
High Energy Physics of the U.S. Department of Energy under Contract
DE-AC03-76SF00098, in part by the National Science Foundation under
grant PHY-95-14797 and in part by the U.S. Department of energy under
contract DOE/ER/01545--700.}

%

\vskip 0.3in

Riccardo Barbieri$^1$,
Lawrence J. Hall$^2$, Stuart Raby$^3$ and Andrea Romanino$^1$

\vskip 0.1in

{{}$^1$ \em Physics Department, University of Pisa\\
     and INFN, Sez.\ di Pisa, I-56126 Pisa, Italy}

\vskip 0.1in

{{}$^2$ \em Department of Physics and
     Lawrence Berkeley National Laboratory\\
     University of California, Berkeley, California 94720, USA}
\vskip 0.1in

{{}$^3$ \em Department of Physics, The Ohio State University\\
     Columbus, Ohio 43210, USA}

\end{center}

\vskip .3in

\begin{abstract}
\noindent
A general operator expansion is presented for quark and lepton mass matrices
in unified theories based on a U(2) flavor symmetry, with breaking
parameter of order $V_{cb} \approx m_s/m_b \approx
\sqrt{m_c/m_t}$. While solving the supersymmetric flavor-changing
problem, a general form for the Yukawa couplings follows, leading to 9
relations among the fermion masses and mixings, 5 of which are
precise. The combination of grand unified and U(2) symmetries provides 
a symmetry understanding for the anomalously small values of $m_u/m_c$ and
$m_c/m_t$. A fit to the fermion mass data leads to a prediction for the angles
of the CKM unitarity triangle, which will allow a significant test of these
unified U(2) theories. A particular SO(10) model provides a simple realization
of the general operator expansion. The lighter generation masses and
the non-trivial structure of the CKM matrix are generated from the exchange of
a single U(2) doublet of heavy vector generations. This model suggests that
CP is spontaneously broken at the unification scale --- in which case there
is a further reduction in the number of free parameters.
\end{abstract}

\end{titlepage}

\section{Flavor in Supersymmetry} \label{sec:flavor}

\subsection{Fermions}

Is it possible for the pattern of fermion masses and mixing angles to
be explained in a qualitative and quantitative way by a suitable
extension of the symmetries of the Standard Model? Despite great effort, the
answer to this fundamental question remains elusive. In this paper we explore
the combined consequences of vertical grand unified symmetries, which lead to
the successful prediction of gauge coupling unification~\cite{georgi:74a}, and
horizontal flavor symmetries, which act on the three unified generations.

The measured values of the bottom quark and the $\tau$
lepton masses are compatible with their equality at a unification
scale~\cite{chanowitz:77a}, establishing a nice consistency with the
heaviness of
the top quark in the case of a supersymmetric
theory~\cite{ibanez:83a}.
On the other hand, the interpretation of the light masses and of the
mixing angles constitutes a formidable challenge.
Several relations have been noticed in the past, sometimes justified
on a theoretical basis, like, e.g.,
$|V_{us}|\simeq(m_d/m_s)^{1/2}$%
~\cite{gatto:68a},
$m_{\mu}\simeq 3 m_s$ and $m_d\simeq 3 m_e$~\cite{georgi:79a}, or
$|V_{ub}/V_{cb}|\simeq(m_u/m_c)^{1/2}$~\cite{harvey:80a}, involving
the masses
and the CKM matrix elements renormalized at the unification scale.
Apparently missing, however, is a coherent overall picture
based on a minimum number of assumptions and capable of experimental
predictions. Ref.~\cite{anderson:94a} is an attempt in this direction.

We seek grand unified and flavor symmetries acting on the three generations,
$\psi$, and spontaneously broken by a set of fields $\phi$,
so that the Yukawa interactions can
be built up as an expansion in $\phi/M$ where $M$ is the cutoff of an effective
theory:

\begin{equation}
[\psi H (1 + {\phi \over M} + ...)\psi]_F
\label{fermass}
\end{equation}
and $H$ contains the Higgs doublets.
This expansion should yield the observed hierarchical structure of the
fermion masses and mixings, hence the leading term in (\ref{fermass}) should
not give masses to the lighter generations nor should it give rise to quark
mixing. The structure of (\ref{fermass}) should be sufficiently constrained
so that there are few relevant parameters and a quantitative fit to the data
should lead to predictions for quantities that can be measured.

\subsection{Scalars}
\label{sec:scalars}

In supersymmetric theories there are mass and interaction matrices for the
squarks and sleptons, leading to a rich flavor structure. In particular,
if fermions and scalars of a given charge have mass matrices which are not
diagonalized by the same rotation, new mixing matrices, $W$, occur at gaugino
vertices. The squark and slepton mass matrices will be constrained by the
grand unified and flavor symmetries and arise dominantly from

\begin{equation}
[\psi^\dagger(1 + {\phi \over M} + {\phi^\dagger \phi \over M^2} + ...)
z^\dagger z \psi]_D.
\label{scalmass}
\end{equation}
where a superfield notation has been used and $z$ is a supersymmetry breaking
spurion, taken dimensionless, with $z = m \theta^2$\footnote{A (possibly
partial) list of papers that have used flavor symmetries to
constrain the generation structure of both the fermion and sfermion
masses is given in  
Ref.~\cite{list}. This is alternative to the view that the Yukawa
coupling and the supersymmetry breaking sectors are decoupled and that
the sfermion masses are almost degenerate, in generation space, due to
some particular dynamical mechanism~\cite{list2}.}. 

Equation (\ref{fermass}) must contain a large interaction which
generates the large top quark Yukawa coupling. Since (\ref{fermass}) and
(\ref{scalmass}) are governed by the same symmetries, it follows that at
least some of the scalars of the third generation are likely to have
masses very 
different from their first and second generation partners: $m^2_3 - m^2_{1,2}
\approx m^2$, where $m^2$ is the average scalar mass
squared~\cite{barbieri:94a}. 
Such a mass splitting does not lead to excessive
flavor-changing effects provided the new mixing matrices have elements
involving the third generation which are no greater than those of the CKM
matrix, $V$: $|W_{3i}| \ltap |V_{3i}|$ and $|W_{i3}| \ltap |V_{i3}|$,
where $i=1,2$~\cite{barbieri:95a}. If the approximate equality holds,
then there are contributions to flavor-changing
processes which are interesting -- we call this the ``1,2---3"
signal~\cite{barbieri:proc}. 

On the contrary, in general there should be considerable degeneracy between the
scalars of the first two generations: $m^2_2- m^2_1 \ll m^2$. For
Cabibbo-like mixing for $W_{12}$, the observed CP violation in the $K$ system
requires~\cite{gabbiani}

\begin{equation}
{m^2_2 - m^2_1 \over m^2} \; \ltap \; {10^{-3} \over \sin \phi} \left( {m
\over 300 \mbox{GeV}} \right)
\label{epsilon}
\end{equation}
in the charge $-1/3$ sector, where $\phi$ is the relevant CP violating phase.
In the lepton sector, the corresponding limit from
$\mu \rightarrow e \gamma$ and $\mu \rightarrow e$ conversion is
$(m^2_2 - m^2_1) / m^2 \; \ltap \; 10^{-2}$ $(m / 100
\mbox{GeV})$~\cite{gabbiani}. 
These constraints we refer to as the ``1---2"
problem~\cite{barbieri:proc}. 

We conclude that the flavor symmetry should yield mixing of the third
generation with the lighter two which is at most CKM-like, giving a
possible ``1,2---3" signal,
while in the light sector it should yield small fermion masses and small scalar
mass splittings, solving the ``1---2" problem. We find that all these features
are satisfied by a U(2) flavor symmetry in which the lighter two generations
transform as a doublet and the third generation as a trivial singlet: $\psi =
\psi_a + \psi_3$, providing the symmetry breaking parameter is governed by
$V_{cb} \approx m_s/m_b \approx \sqrt{m_c/m_t}$.
\footnote{This is to be contrasted with previous attempts to address the issue
of flavor in supersymmetry using a U(2) flavor symmetry, where the symmetry
breaking parameter was taken to be $\sqrt{|V_{cb}|} \approx \sqrt{m_s/m_b}$
\cite{pomarol,barbieri:ref8}.}

\section{U(2) and its Breaking}

In a grand unified theory, for example based on the group SO(10), the maximal
flavor group is U(3), with the three generations transforming as a triplet.
This flavor group will be strongly broken by the large top Yukawa coupling to
U(2), which is the flavor group studied in this paper.
The three generations $\psi$ are taken to transform as $\bf 2\oplus 1$,
\[
\psi=\psi_a\oplus\psi_3.
\]
Taking the Higgs bosons to be flavor singlets, the Yukawa interactions
transform as:
\[
(\psi_3 \psi_3), \qquad  (\psi_3 \psi_a), \qquad  (\psi_a \psi_b).
\]
We assume that the quark and lepton mass matrices can be adequately described
by terms in (1) up to linear order in the fields $f$, ``flavons'', non
trivial under U(2). Hence the only relevant U(2)
representations for the fermion mass matrices are $1$, $\phi^a$, $S^{ab}$ and
$A^{ab}$, where $S$ and $A$ are symmetric and antisymmetric tensors, and the
upper indices denote a U(1) charge opposite to that of $\psi_a$. The
transformation properties of these fields under the unified gauge group is
discussed in the next section. The observed
hierarchy of fermion masses of the three generations leads us to a flavor
symmetry breaking pattern
\begin{equation}
U(2) \stackrel{\epsilon}{\rightarrow} U(1) \stackrel{\epsilon'}{\rightarrow} 0
\label{u2br}
\end{equation}
so that the generation mass hierarhies $m_3/m_2$ and $m_2/m_1$ can be
understood in terms of the two symmetry breaking parameters $\epsilon$ and
$\epsilon'$.

Fermion masses linear in $\epsilon$ can arise only from:
$\vev{\phi^2}/M \equiv \epsilon_\phi$
and $\vev{S^{22}}/M \equiv \epsilon_S$, giving Yukawa matrices of the form
\[
\left( \begin{array}{cc}
\epsilon_S & \epsilon_\phi \\
\epsilon_\phi & 1
\end{array} \right)
\]
in the heavy $2 \times 2$ space. The breaking $\epsilon_\phi$ is necessary to
describe $V_{cb}$, while $\epsilon_S$ is necessary if U(2) alone is to solve a
possible ``1,2---3" flavor-changing problem.
In theories without $\epsilon_S$,
the symmetry breaking parameter $\epsilon$ is of order $\sqrt{m_s/m_b}$ to 
account for the strange quark mass. This leads to excessive contributions to
$\epsilon_K$, unless the mass splitting $m_3^2 - m_{1,2}^2$ is
taken to be  
considerably less than $m^2$, as in~\cite{pomarol,barbieri:ref8}.
On the contrary, the U(2) symmetry breaking parameter $\epsilon\simeq
|V_{cb}|\simeq m_s/m_b$
leads, via (\ref{scalmass}), 
to a scalar mass splitting $(m_2^2 - m_1^2)/m^2 \approx \epsilon^2 \approx 10^{-3}$.
The constraint of (\ref{epsilon}) is just satisfied: the ``1---2" problem has
become an interesting ``1---2" signature.

In principle
there are a variety of options for the last stage of symmetry breaking in
(\ref{u2br}). We assume
\begin{itemize}
\item The theory contains one of each of the fields $\phi^a, S^{ab}$ and
$A^{ab}$.
\item 
The non zero vevs of $\phi^a$, $S^{ab}$ and $A^{ab}$ each participate
in only one stage of the symmetry breaking in~(\ref{u2br}).

\end{itemize}
In the basis where $\phi^2={\cal O}(\epsilon)$ and $\phi^1=0$,
$S^{22}={\cal O}(\epsilon)$ and all other components of $S$ vanish
-- if they were non-zero they would break U(1) at order
$\epsilon$, which is excluded by (\ref{u2br}).
Hence, these assumptions imply that the last stage of the flavor symmetry
breaking of
(\ref{u2br}) must be accomplished by $A^{12}$ of order $\epsilon'$. 

To linear order in $f/M$, the symmetry breaking
pattern (\ref{u2br}) leads to Yukawa matrices in up, down and charged
lepton sectors of the form:
\[
\left( \begin{array}{ccc}
0 & \epsilon' & 0 \\
-\epsilon' & \epsilon & \epsilon \\
0 & \epsilon & 1
\end{array} \right).
\]
Such a pattern agrees qualitatively well with the observed quark and
lepton masses and mixings, with three exceptions:
\begin{itemize}
\item $m_b \approx m_\tau \ll m_t$.

This may not be a puzzle. It could result
as a consequence of $h_1$, the light Higgs which couples to the $D/E$ sectors,
having a small, order $m_b/m_t$, component of the Higgs doublet in the unified
multiplet $H$. Such Higgs mixing could be understood in terms of symmetry
breaking in the Higgs sector, and would reduce the Yukawa matrices
$\lambda^{D/E}$ by a small factor, $\xi$, relative to the Yukawa matrix
$\lambda^U$.

\item $m_c/m_t \ll m_s/m_b, m_\mu/m_\tau$.

\item $m_u m_c/m_t^2 \ll m_d m_s/m_b^2, m_e m_\mu/m_\tau^2$.

\end{itemize}

Neither the U(2) symmetry nor Higgs mixing appears to give a fermion mass
hierarchy which is larger in the $U$ sector than in the $D/E$ sectors.
Hence the central question which this U(2) framework must address is {\em Why
do $\lambda^U_{22}$ and $\lambda^U_{21}$ vanish at order $\epsilon$ and
$\epsilon'$ respectively?}

\section{SU(5) Analysis}
\label{sec:SU(5)}
\subsection{Suppression of $\lambda^U_{22,12}$}

The central issue of why $\lambda^U_{22}$ and $\lambda^U_{21}$ vanish at
order $\epsilon$ and $\epsilon'$, respectively, can be answered using SU(5),
which is contained in all grand unified symmetry groups.
To linear order in the $\phi/M$, the expansion (1) in the case of
SU(5)$\times$U(2) has the form
\begin{equation}
T_3 H T_3 + T_3 \bar{H} \bar{F}_3
\label{su5a}
\end{equation}
\begin{equation}
+{1 \over M} \left(T_3 \phi^a H T_a  + T_3 \phi^a \bar{H} \bar{F}_a
+ \bar{F}_3 \phi^a \bar{H} T_a \right)
\label{su5b}
\end{equation}
\begin{equation}
+{1 \over M} \left(T_a(S^{ab} H + A^{ab} H) T_b + T_a(S^{ab} \bar{H}
+ A^{ab} \bar{H}) \bar{F}_b \right)
\label{su5c}
\end{equation}
where $T$ and $\bar{F}$ are $\bf 10$ and ${\bf \bar{5}}$ representations of
matter and $H$ and $\bar{H}$ are $\bf 5$ and ${\bf \bar{5}}$ representations of
Higgs, necessary for acceptable third generation masses. In general the $\phi,
S$ and $A$ multiplets can transform as any SU(5) representation with zero
fivality and containing one, or more, SM singlets. The interactions of
(\ref{su5a}), (\ref{su5b}) and (\ref{su5c}) 
are understood to include all possible SU(5) invariants. The second operator of
(\ref{su5a}) leads to the well-known SU(5) mass relation 
\cite{chanowitz:77a,ibanez:83a}
\begin{equation}
m_b = m_\tau
\label{b/tau}
\end{equation}
at the unification scale.

The couplings $\lambda^U_{22,12}$ arise from the $T_a T_b$ terms of
(\ref{su5c}), while the couplings $\lambda^{D,E}_{22,12}$ arise from the
$T_a \bar{F}_b$ terms. These terms are distinguished because  $T \times T$
possesses a definite symmetry, $\overline{\bf 5}_s + \overline{\bf 45}_a$ for
components containing a Higgs 
doublet, while  $T \times \bar{F}$ does not.
The vanishing of  $\lambda^U_{22,12}$ at order $\epsilon, \epsilon'$ is
immediate if the SU(5) representations of $S$ and $A$ are such that $SH$
and $AH$ do not transform as $\bf 5$ and $\bf 45$ respectively.
For example, since $A^{ab}H$ is
antisymmetric in flavor, it couples to $T_a T_b$ only if it is
conjugate to the
antisymmetric product of ${\bf 10} \times {\bf 10}$, which is a $\bf
\overline{45}$. 
For $\lambda^{D,E}_{22,12}$ to be non-zero at order $\epsilon, \epsilon'$,
$SH$ and $AH$ must transform as $\bf 45$ and $\bf 5$, or 
the multiplets $S,
A$ must transform as {\bf 75}, {\bf 1} respectively. This implies that
$\lambda^{D,E}_{22}$ arise from $S\bar{H} \sim {\bf \overline{45}}$,
leading to the Georgi-Jarlskog~\cite{georgi:79a} mass relation
\begin{equation}
m_\mu = 3 m_s \left(1 - {m_d \over m_s} \right)
\label{GJ}
\end{equation}
at the unification scale. Similarly,
$\lambda^{D,E}_{12}$ arise from $A\bar{H} \sim {\bf \bar{5}}$, leading to the
highly successful determinantal mass relation
\begin{equation}
m_s m_d = m_\mu m_e
\label{DET}
\end{equation}
at the unification scale.
In any grand unified theory where the flavor symmetry U(2) completely solves the
supersymmetric flavor-changing problem,
SU(5) provides a symmetry understanding for the vanishing of
$\lambda^U_{22,12}$ at order $\epsilon, \epsilon'$,
and leads to the Georgi-Jarlskog relation (\ref{GJ}) and
the determinantal relation (\ref{DET}) as direct, necessary consequences.

\subsection{Higher order origin for $\lambda^U_{22,12}$}

The SU(5) theory of the previous subsection, described by (\ref{su5a}),
(\ref{su5b}) and  (\ref{su5c}), qualitatively accounts for all fermion masses
and mixings, with the exception that $m_u = 0$, which is a consequence of the
SU(5) and flavor symmetries leading to $T_a A^{ab} H T_b = 0$.
For $m_u$ to be non-zero
at higher order in $\phi/M$, additional fields $\phi$ must be added. We
choose to do this by introducing a field $\Sigma_Y$ which is a trivial flavor
singlet and an SU(5) $\bf 24$. 
The subscript $Y$ is then to recall that the vev $\vev{\Sigma_Y}$ has to
break SU(5), so that it points in the hypercharge direction $Y$.
The observed value for $m_u$ leads to
$\Sigma_Y/M\equiv\rho \approx 0.02$, hence we need only keep terms in
the expansion at 
order $(\phi/M)^2$ which give leading contributions to the masses. These
relevant terms are
\begin{equation}
{1 \over M^2} ( T_a \phi^a \phi^b  H T_b + T_a S^{ab} \Sigma_Y H T_b
+ T_a A^{ab} \Sigma_Y H T_b)
\label{su5d}
\end{equation}
The first operator gives an order $\epsilon^2$ contribution to $m_c/m_t$,
augmenting a contribution of the same order which arises from the
diagonalization of the $U$ mass matrix in the heavy 23 sector.
The second and third operators
lead to contributions to $\lambda^U_{22,12}$ at order $\epsilon \rho$ and
$\epsilon' \rho$ respectively.

\subsection{General Consequences}
\label{sec:general}

The Yukawa matrices which follow from this expansion in SU(5) and U(2)
breaking, via the operators of (\ref{su5a}), (\ref{su5b}), (\ref{su5c}) and
(\ref{su5d}), are
\begin{equation}
\lambda^U = \left( \begin{array}{ccc}
0 & \epsilon' \rho & 0 \\
- \epsilon' \rho & \approx \epsilon \rho&  \approx \epsilon \\
0 &  \approx \epsilon & 1
\end{array} \right)\lambda
\label{su5yuku}
\end{equation}
\begin{equation}
\lambda^{D,E} = \left( \begin{array}{ccc}
0 & \epsilon' & 0 \\
-\epsilon' & (1, -3)\epsilon &  \approx \epsilon \\
0 &  \approx \epsilon & 1
\end{array} \right)\xi
\label{su5yukd}
\end{equation}
where ``$ \approx$"
represents unknown couplings of order unity%
\footnote{These may differ in $D$
and $E$ sectors. If $\phi^a$ is non-singlet under SU(5), then the 23 and 32
entries are all 
arbitrary; whereas if it is singlet there are only three independent Yukawa
couplings describing these entries.}, and $\xi \ll \lambda$ follows from
Higgs mixing, if the light Higgs doublet $\bar{h}$ contains only a small
part
of the doublet in the SU(5) multiplet $\bar{H}$. Yukawa matrices of this
form can be diagonalized perturbatively to give a CKM
matrix~\cite{hall:93a} 
\begin{equation}
V = \left( \begin{array}{ccc}
1                             & s_{12}^D - s_{12}^U e^{i \phi} & -s_{12}^U s \\
s_{12}^U -s_{12}^D e^{i \phi} & e^{i \phi}                     &  s \\
s_{12}^D s                    &  -s                            & 1
\end{array} \right)
\label{CKM}
\end{equation}
where
\begin{equation}
s_{12}^D = \left| {V_{td} \over V_{ts}} \right| =
\sqrt{{m_d \over m_s}} \left( 1 - {m_d \over 2 m_s} \right)
\label{12D}
\end{equation}
and
\begin{equation}
s_{12}^U =  \left| {V_{ub} \over V_{cb}} \right| =
\sqrt{{m_u \over m_c}}.
\label{12U}
\end{equation}
with $m_u$ and $m_c$, as $m_d$ and $m_s$, renormalized at the same scale.

These Yukawa matrices
lead to the qualitative results shown in Tables~\ref{tab:qualitative1}
and~\ref{tab:qualitative2}.  
Since $\rho \simeq \epsilon \simeq \xi \simeq 0.02$ and $\epsilon'
\simeq 0.004$, the cutoff scale, $M$,
is 30---50 times larger than the unification scale,
at which SU(5) breaks and U(2) is broken to U(1). The scale of flavor U(1)
symmetry breaking is about an order of magnitude beneath the
unification scale. 

Since the 13 flavor observables are given in terms of 4 small parameters, the
9 approximate predictions can be taken to be $m_t, m_b, m_c, m_s, m_d$ from
Table~\ref{tab:qualitative1} and the 4 CKM parameters from
Table~\ref{tab:qualitative2}. 
\begin{table}
\begin{center}
\vskip 0.25in
\large
\renewcommand{\arraystretch}{1.2}
\begin{tabular}{|||c|c||c|c||c|c||c|c|||}
\hline
    $m_t$   
  & $1$     
  & $ {m_c \over m_t}$      
  & $\epsilon^2$ 
  & $ {m_u \over m_c}  $
  & $ {\epsilon'^2 \over \epsilon^2}{\rho^2\over\epsilon^2}$  
  & $ {m_u m_c \over m_t^2}$ & $\epsilon'^2 \rho^2$\\
\hline
    $m_b$  
  & $\xi$ 
  & $ {m_s \over m_b}$      
  & $\epsilon$   
  & $ {m_d \over m_s}  $
  & $ {\epsilon'^2 \over \epsilon^2}$ 
  & $ {m_d m_s \over m_b^2}$ & $\epsilon'^2$\\
\hline
    $m_\tau$ 
  & $\xi$ 
  & $ {m_\mu \over m_\tau}$ 
  & $3 \epsilon$ &  $ {m_e \over m_\mu}$
  & $ {1 \over 9}{\epsilon'^2 \over \epsilon^2}$ 
  & $ {m_e m_\mu \over m_\tau^2}$
  & $\epsilon'^2$\\
\hline
\end{tabular}
\normalsize
\end{center}
\caption{\em Qualitative predictions for quark and lepton masses in
unified {\rm U(2)} theories}
\label{tab:qualitative1}
\end{table}
\begin{table}
\begin{center}
\vskip 0.25in
\large
\renewcommand{\arraystretch}{1.2}
\begin{tabular}{||c|c|c|c||}
\hline
    $|V_{cb}|$  
  & $ {|V_{td}| \over |V_{ts}|}$ 
  & $ {|V_{ub}| \over |V_{cb}|}$ 
  & $\phi$ \\
\hline
    $\epsilon$  
  & $ {\epsilon' \over \epsilon}$ 
  & $ {\epsilon' \over \epsilon} {\rho \over \epsilon}$ 
  & $1$ \\
\hline
\end{tabular}
\normalsize
\end{center}
\caption{\em Qualitative predictions for {\rm CKM} matrix elements in unified
{\rm U(2)} theories}
\label{tab:qualitative2}
\end{table}
Of these 9 approximate relations, it is straightforward to see that 5 are in
fact precise, having no dependence on the unknown coefficients labelled by
``$\approx$". Three of these are mass
relations between the $D$ and $E$ sectors:
the SU(5) $m_b/m_\tau$ relation of (\ref{b/tau}),
the Georgi-Jarlskog relation of (\ref{GJ}),
and the determinantal relation of (\ref{DET}).

These mass relations are corrected by higher dimension operators involving the
$\Sigma_Y$ field, leading to uncertainties of 2---3\%. In addition,
Eq.~(\ref{GJ})
receives a correction of relative order $\epsilon$ from the
diagonalization of the $D/E$-mass matrices in the heavy 23 sector. These mass 
relations are also corrected by loops at the weak scale with internal
superpartners, as are all entries in the Yukawa matrices (\ref{su5yuku}) and
(\ref{su5yukd}). For $\tan \beta
\leq 3$, we estimate these corrections to be less than 2\%, whereas, for large
$\tan \beta$, they can be significantly larger~\cite{blazek:95a}.

The final two precise relations are those of (\ref{12D}) and (\ref{12U}).
These follow purely from the zero entries of  (\ref{su5yuku}) and
(\ref{su5yukd}), together with the antisymmetry of the 12 entries, and an
 approximate perturbative diagonalization of the Yukawa matrices. The zeros and
antisymmetry of the 12 entries are upset by higher dimension operators only
if additional U(2) breaking fields are present.
The approximate diagonalization means that these relations are corrected at
order $\epsilon$ and $\epsilon^2/\rho$ respectively.

The unified U(2) scheme described above provides a simple symmetry
framework leading to the patterns of Tables~\ref{tab:qualitative1}
and~\ref{tab:qualitative2}, and requiring
the 5 precise relations of (\ref{b/tau}),
(\ref{GJ}), (\ref{DET}), (\ref{12D}) and  (\ref{12U}). In this section we have
used SU(5) as the unified symmetry, as it is sufficient to reach our
conclusions; this SU(5) theory may be embedded in a unified U(2) theory
with larger gauge group.

\subsection{The $Q$ Problem}

Each of the precise relations of the previous subsection,
(\ref{b/tau}), (\ref{GJ}), (\ref{DET}), (\ref{12D}) and  (\ref{12U}), are
apparently in good agreement with the data. With the exception of
(\ref{b/tau}), which receives large radiative corrections from the top Yukawa
coupling, these relations involve at least one quantity which is not known from
experiment to better than 20\% or more. However, there is a combination
of these 
quantities which has been determined, using second order chiral perturbation
theory for the pseudoscalar meson masses, to 3.5\% accuracy
\begin{equation}
Q = {{m_s \over m_d} \over \sqrt{1- {m_u^2\over m_d^2}}} = 22.7 \pm 0.08 
\label{Q}
\end{equation}
with a possible ambiguity related to an experimental discrepancy concerning the
$\eta \rightarrow \gamma \gamma$ decay~\cite{leutwyler}.

We find this value for $Q$ conflicts with
the precise relations of the previous subsection.
Combining  (\ref{GJ}) and (\ref{DET}) leads to a determination of
\begin{equation}
{m_s \over m_d} = {1 \over 9} { m_\mu \over m_e} \left(1 + 18 {m_e \over m_\mu}
\right) = 25,
\label{s/d}
\end{equation}
implying that $Q$ is larger than 25 by an amount that depends on $m_u/m_d$.
Using (\ref{b/tau}), (\ref{DET}), (\ref{12U}), but not (\ref{GJ}), one finds
\begin{equation}
\left| {V_{ub} \over V_{cb}} \right| =
\sqrt{{m_u \over m_d}}
\left({ 1 \over 1 - {m_u^2 \over m_d^2}} \right)^{{1 \over 8}}
\left({1 \over Q} \right)^{{1 \over 4}}
\left({m_e m_\mu \over  m_\tau^2} \right)^{{1 \over 4}}
\sqrt{{m_b \over m_c}}\sqrt{{\eta_c \over \eta_b}} \sqrt{y_t}\\
= 0.076
\sqrt{{m_u \over m_d}}
\left({ 1 \over 1 - {m_u^2 \over m_d^2}} \right)^{{1 \over 8}}
\label{Vub}
\end{equation}
where $\eta_c, \eta_b$ and $y_t$ are renormalization factors discussed in
section~\ref{sec:model}, and have been evaluated for $\alpha_s(M_Z)
= 0.117$, and we have 
used the running masses $m_c = 1.27$ GeV and $m_b = 4.25$ GeV. The experimental
value for $|V_{ub}/V_{cb}| = 0.08 \pm 0.02$ ensures that $m_u/m_d$ cannot be
too small in the U(2) framework, thus making the prediction for $Q$ even
larger. 

How should this $Q$ problem, which is a feature of many textures, be
overcome in the U(2) framework? We have argued that the precise relations
receive corrections at most of order $\epsilon \approx 0.03$, and yet the
conflict between (\ref{Q}) and (\ref{s/d},\ref{Vub}) requires that
(\ref{s/d}) be modified 
by 20\%. We believe that the most probably resolution of this puzzle is the
order $\epsilon$ terms in the 23 and 32 entries of the $D$ and $E$
Yukawa matrices. 
Suppose
they have a size $c \epsilon$, where $c$ is a number of order unity. The 23
diagonalization then leads to corrections of the Georgi-Jarlskog relation
(\ref{GJ}) which can be about $2c \epsilon^2$. In
combining (\ref{GJ}) and (\ref{DET}) 
to obtain the relation (\ref{s/d}) for $m_s/m_d$, one must square (\ref{GJ}),
hence the corrections to (\ref{s/d}) are of order $4c \epsilon^2$,
so that 
$c\approx 2 $ is quite sufficient to resolve the discrepancy. We shall
come back to this point in section~\ref{sec:correcting}.

\section{SO(10) Analysis}
\label{sec:SO(10)}
\subsection{Preliminaries}

The rest of this paper concerns unified theories based on the gauge group
SO(10). This is the smallest group which leads to a unified generation, and
it gives theories of flavor which are considerably more constrained than those
based on SU(5).
The three generations $\psi$ are taken to transform as
($\bf 16$, $\bf 2\oplus 1$): $\psi=\psi_a\oplus\psi_3$. To linear order in the
$\phi/M$ expansion, (\ref{su5a}), (\ref{su5b}) and (\ref{su5c}) are
replaced by\footnote{We do not discuss neutrino masses in this
paper. They involve, in the right handed neutrino mass sector, a
different set of operators from the charged fermion mass sector.} 
\begin{equation}
\psi_3 H \psi_3
\label{so10a}
\end{equation}
\begin{equation}
+{1 \over M} \psi_3 \phi^a H \psi_a
\label{so10b}
\end{equation}
\begin{equation}
+{1 \over M} \psi_a(S^{ab} H + A^{ab} H) \psi_b
\label{so10c}
\end{equation}
where the Higgs doublets which couple to matter are taken to transform as
($\bf 10$, $\bf 1$) under SO(10)$\times$U(2). The fields $\phi^a,
S^{ab}$ and $A^{ab}$ could transform as $\bf 1$, $\bf 45$, $\bf 54$ or
$\bf 210$ 
under SO(10), 
and, for any particular choice, each of the above
operators are taken to include all SO(10) invariants. By comparing
(\ref{so10a}), (\ref{so10b}) and (\ref{so10c}) with
(\ref{su5a}), (\ref{su5b}) and (\ref{su5c}) one finds that SO(10) leads to
a reduction
in the number of operators by a factor of 2, 3 and 2, respectively.
For (\ref{so10a}) this is not significant because the effect of the reduction
is negated by Higgs mixing: for the third generation, the interaction
(\ref{so10a}) together with Higgs
mixing, leads only to the relation $m_b = m_\tau$, as in SU(5).
However, for both (\ref{so10b}) and (\ref{so10c})
the reduction can have significant consequences for fermion masses.

If  $\phi^a$ transforms as an
adjoint of SO(10), (\ref{so10b})
contains three SO(10) invariant operators according to which field
the adjoint is taken to act on. For example, (\ref{so10b}) can be expanded as
\begin{equation}
{1 \over M} \left[ (\psi_3 \phi^a) H \psi_a + \psi_3 (\phi^a H) \psi_a
+ \psi_3  H (\phi^a \psi_a) \right]
\label{so10bb}
\end{equation}
where the parentheses show the action of the adjoint $\phi$.
In fact, of these three invariants only two are independent. 
The adjoint vev gives the quantum number of the field it
acts on. Since the quantum number of the Higgs doublet must be just the
negative sum
of the corresponding quantum numbers of the two matter fermions there
is no loss 
of generality in dropping one of the operators.
If $S$ or $A$ transforms as an adjoint, then the flavor symmetry reduces the
two independent SO(10) invariants to a single one, so (\ref{so10c}) becomes
\begin{equation}
{1 \over M} \psi_a(\{S^{ab},H\} + [A^{ab},H]) \psi_b
\label{so10cc}
\end{equation}
with the understanding that the adjoint acts on the matter {\bf 16}
next to it. 

If $\phi^a$, $S^{ab}$ or $A^{ab}$ transforms as an adjoint, there is in
general a complex parameter, $\kappa$ (or equivalently $\kappa'$),
which descibes the orientation of the vev in group space:
\begin{equation}
X + \kappa Y \mbox{ or } (B-L) + \kappa' T_{3R}
\label{kappa}
\end{equation}
where $X$ is the U(1) generator not contained in SU(5), $Y$ is
hypercharge, $B-L$ is baryon number minus lepton number and $T_{3R}$ is the
neutral generator of SU(2)$_R$.
In this paper we do not discuss the superpotential interactions which determine
the vacuum. Simple models can lead to vevs which point
precisely in the $X, Y, (B-L)$ or $T_{3R}$ directions~\cite{lucas:96a}.
As in the SU(5) case, one will also have to include possibly relevant
terms of order $1/M^2$ as in~(\ref{su5d}).

\subsection{The direct SU(5) extension}
\label{sec:direct}

The SU(5) theories discussed in the previous section can be obtained in a
straightforward way from SO(10) by promoting, for example,
\begin{equation}
S^{ab}({\bf 75}) \rightarrow S^{ab}({\bf 210})
\label{Sab}
\end{equation}
\begin{equation}
A^{ab}({\bf 1}) \rightarrow A^{ab}({\bf 1},{\bf 45})
\label{Aab}
\end{equation}
\begin{equation}
\phi^a({\bf 1},{\bf 24}) \rightarrow \phi^a({\bf 45})
\label{phia}
\end{equation}
\begin{equation}
\Sigma_Y({\bf 24}) \rightarrow \Sigma_Y({\bf 45})
\label{A}
\end{equation}
with vevs taken to point in the same direction in group space as in the SU(5)
theory.

In both theories, $m_{s,\mu}$ $(m_{d,e})$ comes from a single operator at order
$\epsilon$ $(\epsilon')$, and $\lambda^U_{22}$ $(\lambda^U_{12})$
arises from a different operator at order $\epsilon \rho$
$(\epsilon' \rho)$. Hence, these entries do not lead to any up-down
relations. The only difference is that the SO(10)
version of the theory involves a factor 3 reduction in
the number of independent couplings linear in $\phi^a$.
The vev $\vev{\phi^a}$ gives $V_{cb}$ and must
lead to sizable corrections to $m_s$ and $m_\mu$ to solve the $Q$ problem.
A non-zero value for $V_{cb}$ requires that
$\phi^a$ be an SO(10) adjoint rather than singlet, so that, in general, the
$\vev{\phi^a}/M$ contributions to the Yukawa matrices are described by
three complex parameters, the two
independent couplings of (\ref{so10bb}) and $\kappa$. However, non-trivial
predictions could result if there is a reduction in the number of free
parameters, as discussed in the next subsection.

\subsection{Theories with adjoint $\phi^a$ and $S^{ab}$}
\label{sec:adjoint}

In this subsection we introduce an alternative class of SO(10)
theories, which does not lead to the SU(5) theory of
section~\ref{sec:SU(5)}. 
The fields $\phi^a$ and $S^{ab}$ cannot be SO(10) singlets, since they
would lead 
to unacceptable values for $V_{cb}$ and for $m_{c, s, \mu}$, respectively. In
the rest of this paper, we consider the next simplest case where they are both
SO(10) adjoints, and $A$ is singlet or adjoint.
In general, the orientation of each adjoint vev involves a
complex parameter (\ref{kappa}). However, these are strongly constrained by
phenomenology:
\begin{itemize}
\item
\begin{equation}
V_{cb} \neq 0 \;\; \Longrightarrow \;\; \kappa_\phi' \neq 0
\label{kappaphi}
\end{equation}
Three interesting special cases are $\vev{\phi^a} \propto X, Y, T_{3R}$.

\item
\begin{equation}
\lambda^U_{22} \ll \epsilon \;\; \Longrightarrow \;\; \kappa_S' = 0
\label{kappas}
\end{equation}
This is the only possibility for which (\ref{so10c}) avoids giving $m_c$ at
order $\epsilon$, however
this orientation does not give a contribution to $m_{s,\mu}$ at
order $\epsilon$ either.

\item
\begin{equation}
\lambda^U_{12} \ll \epsilon' \;\; \Longrightarrow \;\; A({\bf 1}) \;\mbox{ or }
\; A({\bf 45}), \; \; \mbox{with} \; \kappa_A = 0
\label{kappaa}
\end{equation}
A({\bf 1}) gives $\lambda^{D/E}_{12}=0$ while A({\bf 45}) gives
$\lambda^{D/E}_{12}= O(\epsilon')$.

\item
\begin{equation}
m_u \neq 0 \;\; \Longrightarrow \;\; \kappa_{\Sigma_Y} \neq 0
\label{kappasig}
\end{equation}
The $\Sigma_Y$ field, necessary for eventual non-zero values of
$\lambda^U_{22,12}$, will only give $m_u \neq 0$ if its vev breaks SU(5).
\end{itemize}

This class of theories clearly requires a new ingredient: {\em What is the
origin of $\lambda^{D/E}_{22}=O(\epsilon)$}?  If $A$ is a singlet, there is
also the need for an origin for $\lambda^{D/E}_{12}=O(\epsilon')$.
These new ingredients must still suppress $\lambda_{22,12}^U$.
These difficulties have arisen because the vevs $\vev{A({\bf 1})}$ and
$\vev{S} \propto B-L$ preserve a $u \leftrightarrow d$ interchange symmetry.
This is a clear indication that SO(10) should be broken to SU(5) at a
mass scale larger than these vevs. This is done most easily by introducing an
adjoint $\Sigma_X$ with $\vev{\Sigma_X} \propto X$ having a magnitude not far
from the cutoff $M$, for example $M/3$. The scales of the vevs in the
classes of theories which we have discussed in this paper are shown in
Figure \ref{fig:scales}.
\begin{figure}
\small
\begin{center}
\setlength{\unitlength}{0.45\textwidth}
\begin{picture}(2.05,0.825)(-1.05,-0.42)
\put(-1.05,-0.41){\framebox(1.0,0.825){}}
\put(-1.05,-0.41){\framebox(2.05,0.825){}}
\thicklines

\put(-0.90,0){\line(1,0){0.35}}
\put(-0.90,0.3){\line(1,0){0.35}}
\put(-0.90,-0.2){\line(1,0){0.35}}
\put(-1,0.3){$M$}
\put(-1,0.0){$M_G$}
\put(-0.51,0){$\vev{S}\simeq\vev{\phi}\simeq\vev{\Sigma_Y}$}
\put(-0.51,-0.20){$\vev{A}$}
\put(-0.51,0.22){\makebox(0.35,0.08)[l]{[${\rm SO}_{10}\times{\rm U}_2$]}}
\put(-0.90,0.22){\makebox(0.35,0.08){${\rm SU}_5\times{\rm U}_2$}}
\put(-0.90,-0.08){\makebox(0.35,0.08){${\rm SU}_{3,2,1}\times{\rm U}_1$}}
\put(-0.90,-0.28){\makebox(0.35,0.08){${\rm SU}_{3,2,1}$}}
\put(-1.05,-0.4){\makebox(1,0.08){(a)}}

\put(0.15,0){\line(1,0){0.35}}
\put(0.15,0.2){\line(1,0){0.35}}
\put(0.15,0.3){\line(1,0){0.35}}
\put(0.15,-0.2){\line(1,0){0.35}}
\put(0.05,0.3){$M$}
\put(0.05,0.2){$M$}
\put(0.05,0.0){$M_G$}
\put(0.0,0.2){\makebox(0.05,0.1)[bl]{$\simeq$}}
\put(0.54,0){$\vev{S}\simeq\vev{\phi}\simeq\vev{\Sigma_Y}$}
\put(0.54,-0.20){$\vev{A}$}
\put(0.54,0.20){$\vev{\Sigma_X}$}
\put(0.15,0.22){\makebox(0.35,0.08){${\rm SO}_{10}\times{\rm U}_2$}}
\put(0.15,0.12){\makebox(0.35,0.08){${\rm SU}_5\times{\rm U}_2$}}
\put(0.15,-0.08){\makebox(0.35,0.08){${\rm SU}_{3,2,1}\times{\rm U}_1$}}
\put(0.15,-0.28){\makebox(0.35,0.08){${\rm SU}_{3,2,1}$}}
\put(-0.05,-0.4){\makebox(1.05,0.08){(b)}}
\normalsize

\end{picture}
\end{center}
\caption{\em Scales of symmetry breaking vevs appropriate to the class of
theories discussed in sect.~\ref{sec:SU(5)} or in
subsect.~\ref{sec:direct} {\rm (a)} and in
subsect.~\ref{sec:adjoint} {\rm (b)} respectively}
\label{fig:scales}
\end{figure}

The terms in (\ref{so10a},\ref{so10b},\ref{so10c})
should be replaced with
\begin{equation}
\psi_3 f_1 \left( {\Sigma_X \over M} \right) H \psi_3
\label{so10aaa}
\end{equation}
\begin{equation}
+{1 \over M} \psi_3 \phi^a f_2 \left( {\Sigma_X \over M} \right) H \psi_a
\label{so10bbb}
\end{equation}
\begin{equation}
+{1 \over M} \psi_a \left(S^{ab} f_3 \left( {\Sigma_X \over M} \right)
+ A^{ab} f_4 \left( {\Sigma_X \over M} \right) \right)H \psi_b
\label{so10ccc}
\end{equation}
where the functions $f_i$ contain terms to all orders in $\Sigma_X/M$, and each
term represents all possible SO(10) invariant contractions.

This generalization of the theory implies that $\kappa'_\phi \neq 0$ of
(\ref{kappaphi}) is no longer required. However,
(\ref{kappas},\ref{kappaa},\ref{kappasig}) are still required, and {\em
the orientation of the $S$ and $A$ vevs necessary for the
vanishing of $\lambda^U_{22,12}$ at order $\epsilon, \epsilon'$ leads to the
Georgi-Jarlskog {\rm (\ref{GJ})} and determinantal {\rm (\ref{DET})}
mass relations, respectively.} We
stress that this is not the same group theory which gives the Georgi-Jarlskog
relation in SU(5). The vev $\vev{S^{22}} \propto B-L$ corresponds to a fixed
linear combination of vevs of an SU(5) $\bf 1$ and $\bf 24$ which would occur
only as an accident in SU(5). In fact the $u:d:e$ Clebsch ratios are
different --
$0:1:3$ from $(B-L) f(X)$ in SO(10),
and $0:1:-3$ for the vev of a $\bf 45$ in SU(5).
We note that $\vev{S^{22}} \propto B-L$ can occur even if the dominant breaking
of SO(10) is via $\vev{\Sigma_X}$ to SU(5).
Also this vev in the $B-L$ direction is useful for understanding why the Higgs
doublets could have escaped acquiring masses at the unification scale
\cite{cahn:82a}. 

Non-zero values for $\lambda^U_{22,12}$ at order $\epsilon \rho,
\epsilon' \rho$ are generated by
\begin{equation}
+{1 \over M^2} \psi_a \left(S^{ab} \Sigma_Y f_5 \left( {\Sigma_X \over M}
\right) 
+ A^{ab} \Sigma_Y f_6 \left( {\Sigma_X \over M} \right) \right)H \psi_b
\label{so10ddd}
\end{equation}

\subsection{Yukawa Matrices}

We have discussed U(2) theories of flavor based on SU(5) and SO(10).
{\em All} these
theories lead to the qualitative pattern of quark and lepton masses and
mixings shown in Tables 1 and 2. They all possess 9 approximate mass relations
of which 5 are precise as discussed in section (3.3). The Yukawa matrices for
these theories can be written in the form
\begin{equation}
\lambda^U = \left( \begin{array}{ccc}
0        & \epsilon' \rho  & 0 \\
- \epsilon' \rho & \epsilon \rho' & x_u \epsilon\\
0        & y_u \epsilon   & 1
\end{array} \right)\lambda
\label{u}
\end{equation}

\begin{equation}
\lambda^{(D,E)} = \left( \begin{array}{ccc}
0  &  \epsilon'          & 0 \\
-\epsilon' & (1, \pm3) \epsilon &  (x_d, x_e)\epsilon \\
0  & (y_d, y_e) \epsilon  & 1
\end{array} \right)\xi 
\label{d/e}
\end{equation}
where $x_i, y_i = {\cal O}(1)$. 
All parameters are in general complex, although the
phases of $\lambda$, $\xi$, $\epsilon'$ and the common phase of $x_i$,
$y_i$, 
relative to that of $\epsilon$, do not affect
the quark and lepton masses and mixings. 
In the 22 entry, +3 ($-3$) corresponds to the group theory of $B-L$ (the
{\bf 45} of SU(5)).

Different theories in this class are largely distinguished by the restrictions
placed on the 23 entries, which are not constrained in the general case.
The structure of the other entries is remarkably rigid, and is determined by
just 6 parameters.
 
Diagonalization of these matrices leads to expressions, before standard RG
scalings~\cite{barger:93a} from high to low energies for
\begin{enumerate}
\item
the 6 light masses, relative to the heavy ones
\begin{eqnarray}
\label{1E}
\frac{m_em_\mu}{m_\tau^2}& = & {\epsilon'}^2\\
\label{det}
\frac{m_sm_d}{m_b^2}     & = & \frac{m_em_\mu}{m_\tau^2}\\
\frac{m_um_c}{m_t^2}\frac{m_\tau^2}{m_em_\mu} &=& \rho^2\\
\label{c/t}
\frac{m_c}{m_t}  &=& \epsilon\left|\rho'
e^{i(\hat{\alpha}'-\hat{\gamma}-\theta_u-\phi_u)}-x_u y_u \epsilon\right|\\
\label{mu/tau}
\frac{m_\mu}{m_\tau}  &=& \epsilon\left|\pm3
e^{-i(\hat{\gamma}+\theta_e +\phi_e)}-x_e y_e \epsilon\right|\\
\label{s/b}
\frac{m_s}{m_b}\left(1-\frac{m_d}{m_s}\right)  &=& \epsilon\left|
e^{-i(\hat{\gamma}+\theta_d+\phi_d)}-x_d y_d\epsilon\right|
\end{eqnarray}
where $\rho \rightarrow \rho e^{i\hat{\alpha}}$, $\rho' \rightarrow \rho'
e^{i\hat{\alpha}'}$, $\epsilon \rightarrow \epsilon
e^{i\hat{\gamma}}$, $x_i \rightarrow x_i e^{i 
\theta_i}$ and $y_i \rightarrow y_i e^{i\phi_i}$ with $\rho$, $\rho'$,
$\epsilon$, and  $x_i, y_i$ now real. In view of the Q problem, the
$\epsilon^2$ terms have been kept in~(\ref{mu/tau}) and~(\ref{s/b}),
even though they are non-leading order. 

\item
the $V_{\rm CKM}$ matrix~(\ref{CKM}) with~(\ref{12D}), (\ref{12U}) and
\begin{eqnarray}
\label{s}
s &=& \epsilon|x_de^{-i\theta_d}-x_ue^{-i\theta_u}|\\
\label{phi}
\phi &\simeq& \pi-(\hat{\alpha}-\hat{\gamma}-\phi_u - \theta_u).
\end{eqnarray}
\end{enumerate}
The fermion masses and mixings therefore depend on 9 independent
parameters:
\begin{itemize}
\item
$\lambda$, $\xi$ for the third generation;
\item
5 combinations of ($\rho'$, $\hat{\alpha}'$, $\hat{\alpha}$, $\epsilon$,
$\hat{\gamma}$; $x_i$, $\theta_i$, $y_i$, $\phi_i$) for $m_c/m_t$,
$m_s/m_b$, $m_\mu/m_\tau$, $V_{cb}$ and $\phi$;
\item
$\epsilon'$ and $\rho$ for the first generation masses.
\end{itemize}
This leads to 4 precise predictions\footnote{If the $\epsilon$ correction
terms in~(\ref{mu/tau}) and~(\ref{s/b}) are neglected, there are only 8
independent parameters and the Georgi-Jarlskog relation is recovered as
the 5th precise relation. However, in view of the $Q$ problem, the 9th
parameter is needed.}.

Particular theories of this type will be distinguished by the
values for the parameters $x_i, y_i$ of the 23 and 32 entries. In
general SU(5) 
theories, these entries depend on 6 complex parameters, whereas, in general
SO(10) theories, there are only 3 complex parameters. 
Further predictivity will be possible if
\begin{itemize}
\item $\vev{\phi}$ lies in the $X, Y, B-L$ or $T_{3R}$ directions,
\item CP is spontaneously broken in the sector which involves the lightest
generation, making the three relevant parameters real,
\item the operators are generated by the Froggatt-Nielsen
mechanism~\cite{froggatt:79a}, 
as this produces particular SO(10) contractions. For example,
the three operators of (\ref{so10bb}) are generated by the exchange of the
heavy states ({\bf 16},{\bf 2}),  ({\bf 144},{\bf 1})$\oplus$({\bf
144},{\bf 2})  and ({\bf 16},{\bf 1}), 
respectively.
\end{itemize}

In section~\ref{sec:model} we discuss a simple SO(10) Froggatt-Nielsen model
in which $\rho'=\rho$, $\hat{\alpha}'=\hat{\alpha}$ and the set ($x_i$,
$\theta_i$, $y_i$, $\phi_i$) is reduced to three parameters. Even though
the theory still depends on 9 independent parameters, the form of the
Yukawa matrices leads to a somewhat more constrained fit to the fermion
mass data. In this model a form for the phases can be chosen, which might
arise in a theory with spontaneous CP violation, such that the number of
independent parameters of the Yukawa matrices is reduced to 6, leading to
extremely tight predictions.


\section{Predicting the angles of the CKM unitarity triangle}
\label{sec:predicting}

By means of the precise relations, Eqs.~(\ref{CKM},\ref{12D},\ref{12U}),
which are a pure consequence of the U(2) symmetry and
Eqs.~(\ref{b/tau},\ref{det}), which, on the contrary, follow from the
full SU(5)$\times$U(2) or SO(10)$\times$U(2) symmetry, it is possible to
predict the values of the angles of the CKM unitarity triangle, defined
as usual as 
\begin{eqnsystem}{albega}
\alpha & = & \mbox{arg}\,(-\frac{V_{tb}^*V_{td}^{}}{V_{ub}^*V_{ud}^{}})\\
\beta  & = & \mbox{arg}\,(-\frac{V_{cb}^*V_{cd}^{}}{V_{tb}^*V_{td}^{}})\\
\gamma & = & \mbox{arg}\,(-\frac{V_{ub}^*V_{ud}^{}}{V_{cb}^*V_{cd}^{}}).
\end{eqnsystem}

Given Eq.~(\ref{CKM}), the following approximate expressions hold for
these angles
\begin{eqnsystem}{albega2}
\alpha&=&\phi\\
\beta&=&\mbox{arg}\,(1-\frac{s^U_{12}}{s^D_{12}}e^{-i\phi})\\
\gamma&=&\pi-\alpha-\beta
\end{eqnsystem}
in terms of the CP violating phase appearing in the CKM matrix.

To obtain these angles, one observes
that the sides of the unitarity triangle $|V_{ub}/V_{cb}|$ and
$|V_{td}/V_{ts}|$ can both be expressed, as in Eq.~(\ref{Vub}), as
functions of $m_u/m_d$, for given $m_c$, $m_b$, $Q$ and $\alpha_s(M_Z)$.
In the same way, one can express $|V_{us}|$,
Eq.~(\ref{CKM},\ref{12D},\ref{12U}), in terms of $m_u/m_d$ and the CKM
phase $\phi$. Or, for given $|V_{cb}|$, one can express the CP violating
parameters in $K$ physics, $\epsilon_K$, and the $B_d$-$\overline{B}_d$
mixing mass, $\Delta m_{B_d}$, in terms of $m_u/m_d$ and $\phi$.
A fit of these quantities will then determine a range of values for
$m_u/m_d$ and $\phi$ or, via Eqs.~(\ref{albega2}), for the angles
$\alpha$, $\beta$, $\gamma$.
A full list of physical quantities which include also the ones relevant
to this fit is given in Table~\ref{tab:inputs}~\cite{pdg:96a}. Since the
uncertainties 
in these observables are very different, hereafter we fix the well
measured ones, those without an asterisk in Table~\ref{tab:inputs}, to
their central values and we fit the remaining ones. In this way the
uncertainties are slightly underestimated.

\begin{table}
\centering
\renewcommand{\arraystretch}{1.1}
\vskip 0.25in
\begin{tabular}{||c|c|c||c|c|c||}
\hline
$m_e$ & $0.511$ MeV  & & $|V_{us}|$ & $0.221 \pm 0.002$ & \cr
$m_\mu$ & $105.7$ MeV & & $|V_{cb}|$ & $0.038 \pm 0.004$ &* \cr
$m_\tau$ & $1777$ MeV & & $|V_{ub}/V_{cb}|$ & $0.08 \pm 0.02$ &* \cr
$Q$ & $22.7 \pm 0.8$ & & $|\epsilon_K|$ & $(2.26 \pm 0.02) 10^{-3}$ & \cr
$(m_s)_{\rm 1 GeV}$  & $(175 \pm 55)$ MeV &* & $\Delta
m_{B_d}/\mbox{ps}^{-1}$ & $0.464 \pm 0.018$ & \cr 
$(m_c)_{m_c}$ & $1.27 \pm 0.05$ GeV & & $\sqrt{B} f_B$ & $(200 \pm
40)$ MeV &* \cr 
$(m_b)_{m_b}$ & $4.25 \pm 0.15$ GeV & & $B_K$ & $0.8 \pm 0.2$ &* \cr
$(m_t)_{m_t}$ & $165 \pm 10$ GeV & & $\alpha_s (M_Z)$ & $0.117 \pm
0.006$ &* \cr 
\hline
\end{tabular}
\caption{\em List of the input physical observables}
\label{tab:inputs}
\end{table}

Assuming that both $\epsilon_K$ and $\Delta m_{B_d}$ are fully
accounted for by the usual SM box diagrams, we take for them
\begin{equation}
\epsilon_K= 4.7\times 10^5  \,e^{i\pi/4}B_KJ\,(4.22\times 10^{-5}
-2.36\,\Re J^{ds}_{ut})
\end{equation} 
and
\begin{equation}
\Delta m_{B_d}=5.0\times\displaystyle
10^3\left(\frac{\sqrt{B}f_B}{180\MeV}\right)^2|V_{td}|^2\mbox{ps}^{-1}
\end{equation}
where 
\begin{equation}
J^{\alpha\beta}_{ij}  = 
V^{}_{i\alpha}V^*_{i\beta}V^*_{j\alpha}V^{}_{j\beta}\enspace
{\renewcommand{\arraystretch}{0.5}
\begin{array}{c}
\alpha,\beta=d,s,b\\
i,j=u,c,t
\end{array}}\quad\mbox{ and }\quad 
J = \Im[J^{bd}_{tu}]\simeq s^D_{12}s^U_{12}s^2 s_\phi.
\end{equation}
These expressions for $\epsilon_K$ and $\Delta m_{B_d}$ involve 
the quantities $B_K$ and $\sqrt{B}f_{B}$, which we take as further
observables, ``measured'' on the lattice.

The results of the fits are shown in Table~\ref{tab:bet}.
As mentioned in sect.~\ref{sec:scalars}, both $\epsilon_K$ and 
$\Delta m_{B_d}$ may be affected by superpartner loops at the weak
scale.
For this reason, we have considered both a fit where $\epsilon_K$ and 
$\Delta m_{B_d}$ are included in the inputs (``constrained'') as well as
a fit where they are not (``unconstrained''). In the last case, we
simply calculate, as a result of 
the fit, the expected contributions to $\epsilon_K$ and 
$\Delta m_{B_d}$ from the SM box diagrams.
We find infact that such contributions can deviate from the measured
values of $\epsilon_K$ and $\Delta m_{B_d}$, in absolute magnitude and
for the central values of $B_K$ and $\sqrt{B}f_{B}$ indicated in
Table~\ref{tab:bet}, in a significant way. Notice in particular that in
the ``unconstrained'' fit, namely the one not including $\epsilon_K$ and
$\Delta m_{B_d}$ among the inputs, the sign of $\epsilon_K$ is not
determined.

\begin{table}
\renewcommand{\arraystretch}{1.3}
\[
\begin{array}{||c|c|c|c||}
\hline
            &\mbox{inputs}
            &\mbox{constrained}
            &\mbox{unconstrained} \\
\hline
m_s/\MeV     
            & 175 \pm 55      
            & 153^{+35}_{-22}  
            & 153\pm 35 \\
|V_{cb}|    
            & 0.038 \pm 0.004 
            & 0.039^{+0.0025}_{-0.0015} 
            & 0.038 \pm 0.004 \\
|V_{ub}/V_{cb}| 
            & 0.08 \pm 0.02 
            & 0.075\pm 0.013
            & 0.075\pm 0.016 \\
\epsilon_K\cdot 10^3
            & 2.26
            & 2.26
            & \pm (1.7^{+1.3}_{-0.1}) \\
B_K               
            & 0.8 \pm 0.2 
            & 0.86\pm 0.16 
            & 0.8 \\

\Delta m_{B_d}/\mbox{ps}^{-1}
            & 0.464
            & 0.464
            & 0.37^{+0.14}_{-0.05} \\
\sqrt{B}f_B/\MeV  
            & 200 \pm 40  
            & 178\pm 18
            & 200 \\
\hline
\alpha_s(M_Z)          
            & 0.117 \pm 0.006 
            & 0.118\pm 0.005 
            & 0.118\pm 0.005 \\
\hline
\end{array}
\]
\caption{\em Fit in the unified {\rm U(2)} theories with (``constrained'') or
without (``unconstrained'') inclusion of $\epsilon_K$ and $\Delta
m_{B_d}$ in the inputs}  
\label{tab:bet}
\end{table}

\begin{figure}
\begin{center}
\epsfig{file=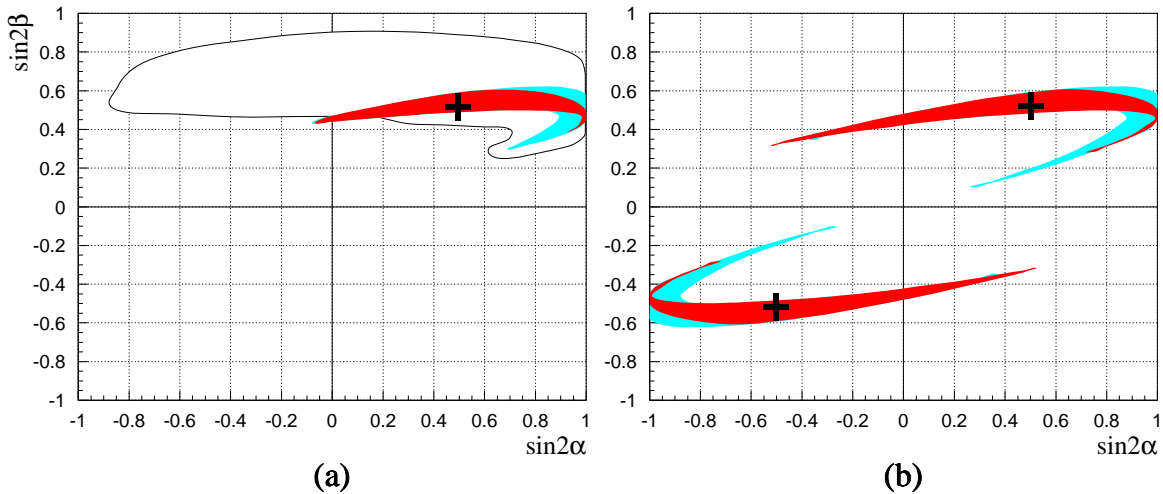,width=\textwidth}
\end{center}
\caption{\em {\rm (a)} 90\% contour plots from the constrained fit
($\epsilon_K$ 
and $\Delta m_{B_d}$ included) for the SM (white area), the unified
{\rm U(2)} theories (light and dark shaded area), the
model of sect.~\ref{sec:model} with free phases (darker area) and with
maximal phases (cross); {\rm (b)} as in Fig.~\ref{fig:sins}a from the
unconstrained fit ($\epsilon_K$ and $\Delta m_{B_d}$ excluded)}
\label{fig:sins}
\end{figure}

As mentioned, these fits allow the prediction of the CKM unitarity
triangle, shown in Figs.~\ref{fig:sins} for the correlation between
$\sin 2\alpha$ and $\sin 2\beta$ at 90\% c.l..
Fig.~\ref{fig:sins}a also includes the current range of values obtained
by doing a fit of the available informations ($|V_{us}|$, $|V_{cb}|$,
$|V_{ub}/V_{cb}|$, $\epsilon_K$, $\Delta m_{B_d}$, $B_K$,
$\sqrt{B}f_{B}$) by a general parametrization of the $V_{\rm CKM}$ matrix in
the SM. No such fit is possible without the inclusion of 
$\epsilon_K$ and $\Delta m_{B_d}$, which explains why the SM range is
not included in Fig.~\ref{fig:sins}b.

At the same time, one obtains $m_u/m_d= 0.76^{+0.10}_{-0.16}$ and
$m_u/m_d= 0.76^{+0.14}_{-0.22}$ in the constrained and unconstrained
fits respectively. These values can be compared with $m_u/m_d=0.553\pm
0.043$, as obtained from chiral perturbation theory and some
supplementary hypothesis~\cite{leutwyler}. 

\section{An explicit SO(10)$\times$U(2) model}
\label{sec:model}
\subsection{Definition and basic formulae}

Within the stated assumptions, everything that has been said so far is
general and is based on an operator analysis. In this section we
describe an  explicit SO(10)$\times$U(2)  model. One purpose for this
is to show that the $Q$ problem can be solved.
This requires a model where all the
corrections of relative order $\epsilon$ to
Eqs.~(\ref{b/tau},\ref{GJ},\ref{DET}) are fully under control. In
turn, this will allow us to detail the numerical fit of the known data
and the predictions for several observables in flavor physics.

We seek a special realization of the superpotential in
Eqs.~(\ref{so10aaa},\ref{so10bbb},\ref{so10ccc},\ref{so10ddd})
generated from a 
renormalizable theory, where the non-renormalizable operators arise
from the exchange of heavy vector-like families (the so called
``Froggatt-Nielsen'' fields). A minimum choice involves one doublet
under U(2), $\chi^a+\overline{\chi}_a$, transforming as
$16+\overline{16}$ under SO(10). The most general, renormalizable,
invariant superpotential involving these vector multiplets, the usual
chiral multiplets, the Higgs ten-plet, the flavon fields $\phi^a({\bf
45})$, 
$S^{ab}({\bf 45})$, $A^{ab}({\bf 1})$ and the adjoint fields $\Sigma_X$ and
$\Sigma_Y$, 
introduced in section~\ref{sec:SO(10)}, is
\begin{equation}
f = \psi_3 H \psi_3 + \chi^a H \psi_a + \overline{\chi}_a(M\chi^a +
\Sigma_X\chi^a + \Sigma_Y\chi^a + \phi^a\psi_3
+ S^{ab}\psi_b + A^{ab}\psi_b) 
\label{Wren}
\end{equation}
where, as usual, dimensionless couplings and SO(10) contractions are
left understood. 

\begin{figure}
\begin{center}
\epsfig{file=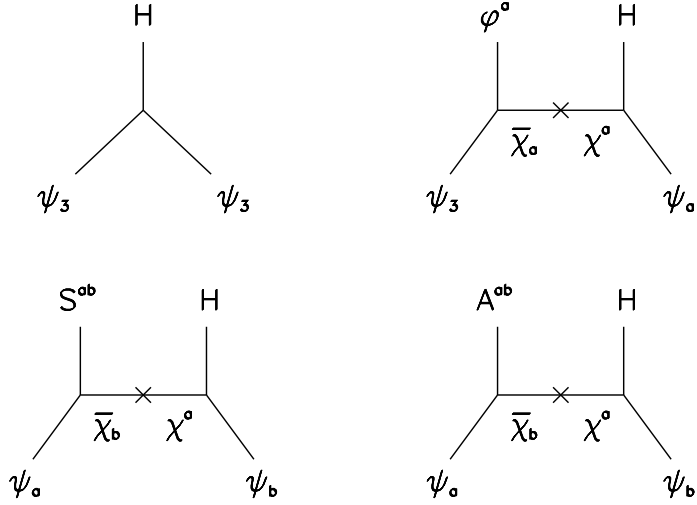,width=0.7\textwidth}
\end{center}
\caption{\em Diagrams generating the Yukawa coupling superpotential in the
{\rm SO(10)$\times$U(2)} model}
\label{fig:diagrams}
\end{figure}

On integrating out the heavy $\chi^a+\overline{\chi}_a$ states, one
generates, from the diagrams shown in Fig.~\ref{fig:diagrams}, a
particular case of the
superpotential~(\ref{so10aaa},\ref{so10bbb},\ref{so10ccc},\ref{so10ddd}).
With 
respect to this general form, the term bilinear in the field $\phi^a$
is absent and only some contractions of the SO(10) indices occur. Also
important is the fact that the
superpotential~(\ref{so10aaa},\ref{so10bbb},\ref{so10ccc},\ref{so10ddd})
contains an 
infinite tower of $(\Sigma_X/M)^n$-operators, which are all under control
in this case. This is
welcome, in view of the fact, already mentioned, that 
$\vev{\Sigma_X}/M$ is not far from unity. In the following
we treat $\vev{\Sigma_X}/M$ exactly and we give explicit
formulae to first order in 
$\vev{\Sigma_Y}/M$, but we control the size of the higher
order terms. 

To be able to write down explicit forms for the Yukawa matrices in
this case, we only need to know the SO(10) properties of the U(2)
doublet $\phi^a$. For the time being we take it to be an SO(10)-adjoint
with its vev point in a generic
SU(3)$\times$SU(2)$\times$U(1)-preserving direction $T$, so that
$\vev{\phi^2}\equiv \vev{\phi}T$.

Following the line of the discussion in the previous section, it is
straighforward to write down explicit expressions for the Yukawa
matrices. After trivial rescalings, one gets
\begin{eqnsystem}{yuk}
\lambda^U &=& \left(\begin{array}{ccc}
0 & \epsilon'\rho & 0\\
-\epsilon'\rho & \epsilon\rho & r T_u\epsilon\\
0 & r T_Q\epsilon & 1
\end{array}\right)\lambda
\\
\lambda^D &=& \left(\begin{array}{ccc}
0 & \epsilon' & 0\\
-\epsilon' & \epsilon & r x T_d\epsilon\\
0 & r T_Q\epsilon & 1
\end{array}\right)\xi
\\
\lambda^E &=& \left(\begin{array}{ccc}
0 & \epsilon' & 0\\
-\epsilon' & 3\epsilon & r x T_L\epsilon\\
0 & r T_e\epsilon & 1
\end{array}\right)\xi
\end{eqnsystem}
where
\begin{eqnsystem}{par}
& \displaystyle r=\frac{\vev{\phi}}{\vev{S}},\qquad
x=\frac{1+\alpha}{1-3\alpha},\qquad
\rho=-\frac{5}{16}\frac{\beta}{\alpha}\frac{1-3\alpha}{1+\alpha},& \\
& \displaystyle\alpha\propto\frac{\vev{\Sigma_X}_u}{M},\qquad
\beta\propto\frac{\vev{\Sigma_Y}_u}{M}&
\end{eqnsystem}
and the normalization of $T$ is immaterial since it can be reabsorbed
in $r$.

These Yukawa matrices have to be compared with the general form given
in Eqs.~(\ref{u},\ref{d/e}). Notice that $\rho=\rho'$ and that all the
coefficients $x_i$, $y_i$ of order unity are now determined by the
parameters $r$ and $x$. In particular, all 
combinations of dimensionless couplings occurring in the diagrams of
Fig.~\ref{fig:diagrams} have been rescaled away. 
By redefining the phases of the matter superfields, it is possible to
make the parameters $\lambda$, $\xi$, $\epsilon'$ and $r$ real and
positive. In general $\rho$, $x$ and $\epsilon$ are complex, so that, from
now on 
\begin{equation}
\rho\rightarrow\rho e^{i\hat{\alpha}},\quad x\rightarrow x
e^{i\hat{\beta}},\quad \epsilon\rightarrow \epsilon e^{i\hat{\gamma}}
\label{phases}
\end{equation}
with $\rho$ real.

\subsection{Solving the $Q$-problem}
\label{sec:correcting}

For any choice of $T$ it is now possible to use
Eq.~(\ref{yuk}) to make a fit of the data. 
Before doing that, let
us discuss again the problem of the corrections to the GJ
relation~(\ref{GJ}).
Notice that Eqs.~(\ref{b/tau}) and (\ref{DET}), which we have used in
sect.~\ref{sec:predicting}, remain unchanged.

As a consequence of the diagonalization of~(\ref{yuk}) in the 23
sector, Eq.~(\ref{GJ}) gets modified as
\begin{equation}
\label{corr}
\frac{m_\mu}{m_\tau} = 3 \frac{m_s}{m_b}\left(1 - {m_d \over m_s}
\right) +\frac{m_\mu}{m_\tau} \Delta
\end{equation}
where $(m_\mu/m_\tau) \Delta$ is an additional contribution that
depends, in 
particular, on the choice of $T$. From Eqs.~(\ref{mu/tau},\ref{s/b}),
after specialization to~(\ref{yuk}), it follows that 
\begin{equation}
|\frac{m_\mu}{m_\tau}\Delta|\leq x\epsilon^2 r^2 
\left|T_e T_L -3 T_d T_Q\right| = 6 x \epsilon^2 r^2
\left|T_u T_Q\right|,
\label{con1}
\end{equation}
where we have used $T_L = -3 T_Q$ and $T_d+T_e = -2 T_u$. From
Eq~(\ref{s}), the parameter $x$ can be bound as 
\begin{equation}
x\left|\frac{T_d}{T_u}\right| \leq 
\left(\frac{|V_{cb}|_G}{|T_u|\epsilon r} +1 \right).
\label{con2}
\end{equation}
while the combination $\epsilon r$ can be obtained from $m_c/m_t|_G$ by
means of Eq~(\ref{c/t}), where the term proportional to $\rho'=\rho$
can be safely neglected for the 
purposes of this discussion. Therefore, from (\ref{con1}), 
\begin{equation}
|\Delta|\leq\Delta_{\rm max}\equiv 6
\left(\frac{m_c}{m_t}\right)_G^{1/2} \left(|V_{cb}|_G
\left|\frac{T_Q}{T_u}\right|^{1/2} +
\left(\frac{m_c}{m_t}\right)_G^{1/2}\right)
\left|\frac{T_u}{T_d}\right|\left/ \frac{m_\mu}{m_\tau}\right.
\label{con3}
\end{equation} 
where the inequality is saturated for maximal phases.  

\begin{figure}
\begin{center}
\epsfig{file=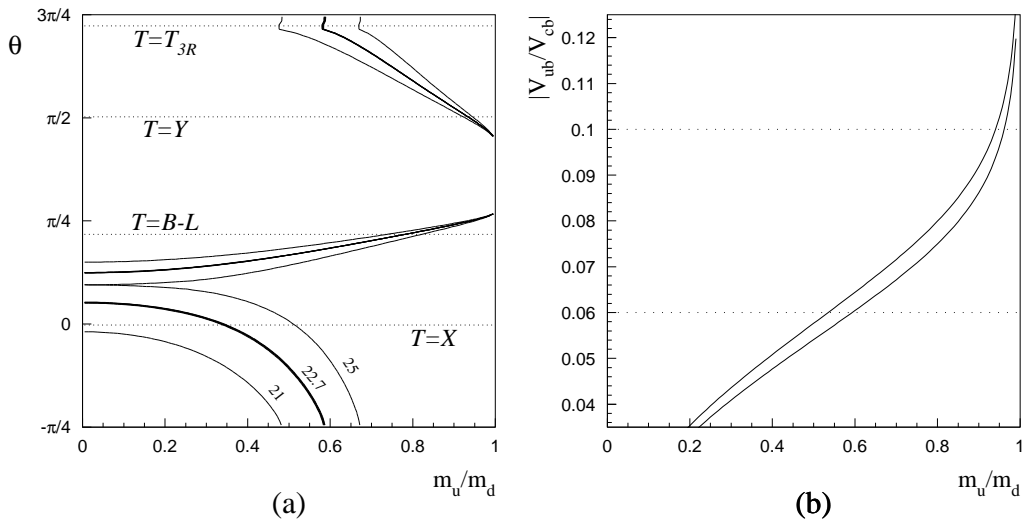,width=0.90\textwidth}
\end{center}
\caption{\em {\rm (a)} contour plot of $Q_{\rm min}$ versus $m_u/m_d$ and
$\theta$ in $T=X\cos\theta+Y\sin\theta$ for central values of $m_c$,
$m_t$, $|V_{cb}|$, $\alpha_s(M_Z)$; (b) $|V_{ub}/V_{cb}|$ versus
$m_u/m_d$ for $\alpha_s(M_Z)=0.117\pm 0.003$ and central values of
$m_b$, $m_c$, $Q$}
\label{fig:con}
\end{figure}

Using~(\ref{corr}) instead of~(\ref{GJ}) gives
\begin{equation}
\label{s/d2}
\frac{m_s}{m_d}=25\cdot(1-2\Delta)
\end{equation}
instead of~(\ref{s/d}), so that, from~(\ref{Q}), 
\begin{equation}
\label{Q2}
Q=25\cdot\frac{1-2\Delta}{\sqrt{1-\frac{m^2_u}{m^2_d}}}
\end{equation}
To see the consistency of this expression with $Q=22.7\pm 0.08$, we
plot in Fig.~\ref{fig:con}a the contours of $Q_{\rm min}\equiv
Q(\Delta=\Delta_{\rm max})$ as function of $m_u/m_d$ and of a
parameter $\theta$ which defines the general superposition of SO(10)
generators: $T=X\cos\theta+Y\sin\theta$. This plot is only weakly
sensitive to the values of $m_c$, $m_t$, $|V_{cb}|$, $\alpha_s(M_Z)$,
then fixed to their central values.
As apparent from this figure, $Q$ as low as $22.7\pm 0.08$ can be
obtained if $T=Y$ or $T=B-L$ for values of $m_u/m_d$ compatible with
$|V_{ub}/V_{cb}|$, plotted in Fig.~\ref{fig:con}b.
This is confirmed and made
explicit by the overall fit discussed in the next subsection. 

\subsection{Parameter fit}

A general fit of the data can be made based on
Eqs.~(\ref{1E}--\ref{phi}), specialized as in~(\ref{yuk}) with
$T={Y}$ or $T={B-L}$. By a
usual analysis, $\xi$ and $\lambda$ are determined by $m_t$ and
$m_\tau$, allowing a prediction for $m_b$ in terms of $\alpha_s$ and
$\tan\beta$.

For the renormalization rescalings~\cite{barger:93a}, we use in particular
\begin{equation}
\frac{m_b}{m_\tau} =
\frac{\eta_b}{\eta_\tau}\prod_a\frac{\zeta^d_a}{\zeta^e_a}\frac{1}{y_t}
\label{RGE}
\end{equation}
where $\eta_b$, $\eta_\tau$ are the scaling factors from the weak
scale to low energy, $\zeta^{d,e}_{a=1,2,3}$ are the gauge couplings
renormalizations from the GUT scale to the weak scale and $y_t$ is the
scaling factor, still from the GUT to the weak scale, due to the top
Yukawa coupling. Eq.~(\ref{RGE}) is appropriate for the low
$\tan\beta$ case, to which we stick in the following. One motivation
for this is to be sure that the weak scale threshold corrections
mentioned in section~\ref{sec:general} do not invalidate the analysis.

The 16 observables in Table~\ref{tab:inputs} depend on 14 parameters: the
10 free flavor parameters, 
the ratio of the two electroweak vevs
$v_2/v_1$, $\alpha_s$, $\sqrt{B} f_B$ and $B_K$, so that the fit has 2
degrees of freedom. Having fixed the more precisely measured quantities
to their central values, the other 6
observables are then fitted by varying the  
4 remaining independent parameters (which we choose to be $\alpha_s$,
$m_u/m_d$, $\cos(\hat{\alpha}-\hat{\gamma})$ and $\cos\hat\beta$).
One should note that the errors in the input observables are mostly
theoretical. 

The results of the fit are shown in Tables~\ref{tab:parameters} and
\ref{tab:fitted} 
respectively for 
the parameters of the model, as defined in
Eqs.~(\ref{yuk},\ref{phases}) with $T={Y}$ and
for the 6 input physical observables whose central values are allowed
to vary. The fit does not determine the relative sign of
$\sin(\hat\alpha-\hat\gamma)$ and $\sin(\hat\beta+\hat\gamma)$, but
this ambiguity does not affect in a significant way any of the
observables listed in Table~\ref{tab:parameters} and \ref{tab:fitted}.
\begin{table}
\renewcommand{\arraystretch}{1.3}
\begin{displaymath}
\begin{array}{||c|c|c||}
\hline
         & \mbox{free phases} 
         & \mbox{max phases}           \\
\hline
\chi^2_{\rm min}/\mbox{d.o.f.} 
         & 0.67/2       
         & 2.1/5    \\
\hline
\epsilon 
         & 0.0162^{+0.0013}_{-0.0008} 
         & 0.0174\pm 0.0002\\
\rho     
         & 0.0201\pm 0.006
         & 0.0205\pm 0.001\\
\epsilon'
         & 0.00414            
         & 0.00414      \\
r(T_u T_Q)^{1/2} 
         & 1.95^{+0.31}_{-0.22} 
         & 2.10\pm 0.07\\
x        
         & 2.56^{+0.4}_{-1.1} 
         & 1.20\pm 0.035\\
\hline
\cos(\hat{\alpha}-\hat{\gamma}) 
         & 0.22^{+0.19}_{-0.33} 
         & 0                           \\
\cos(\hat{\beta} +\hat{\gamma}) 
         & -0.95^{+0.55}_{-0} 
         & -1                    \\
\cos\hat{\beta}                  
         & 0.96^{+0.04}_{-0.06} 
         & +1                    \\
\hline
\end{array}
\end{displaymath}
\caption{\em Parameters of the model, as determined from the fit, for
$T={Y}$ with free phases or maximal phases} 
\label{tab:parameters}
\end{table}
\begin{table}
\renewcommand{\arraystretch}{1.3}
\[
\begin{array}{||c|c|c|c||}
\hline
            &\mbox{inputs}
            &\mbox{free}
            &\mbox{max}                   \\
\hline
m_s/\MeV     
            & 175 \pm 55      
            & 158\pm 28
            & 155\pm 6\\ 
|V_{cb}|    
            & 0.038 \pm 0.004 
            & 0.0391\pm 0.0025
            & 0.0407\pm 0.002\\ 
|V_{ub}/V_{cb}| 
            & 0.08 \pm 0.02 
            & 0.075^{+0.003}_{-0.012}
            & 0.0611\pm 0.001\\
\sqrt{B}f_B/\MeV  
            & 200 \pm 40  
            & 179^{+14}_{-10} 
            & 187\pm 8.5\\
B_K               
            & 0.8 \pm 0.2 
            & 0.84^{+0.18}_{-0.14} 
            & 0.91\pm 0.15\\
\alpha_s          
            & 0.117 \pm 0.006 
            & 0.119\pm 0.005
            & 0.114\pm 0.001\\
\hline
\end{array}
\]
\caption{\em Results of the fit for $T={Y}$ with
free phases or maximal phases} 
\label{tab:fitted}
\end{table}
The fit with $T={B-L}$ gives results which are
all within the uncertainties quoted in Table~\ref{tab:parameters},
\ref{tab:fitted}.

As apparent from Table~\ref{tab:parameters}, all the values of the
phases $\hat{\alpha}$, $\hat{\beta}$, $\hat{\gamma}$ are compatible
with being maximal.
At least for  $\hat{\beta}$ and $\hat{\gamma}$ this is clearly
indicated by the fit itself and it is suggestive of spontaneous CP
violation. 
With this in mind, we have made a fit with all phases fixed at maximal
values, $\hat{\alpha}=\pi/2$, $\hat{\beta}=0$, $\hat{\gamma}=\pi$ for
$T=Y$.
In this case, having still fixed all the inputs without an asterisk in
Table~\ref{tab:inputs} at their central values, only $\alpha_s$ remains
as free parameter to fit the six observables in
Table~\ref{tab:fitted}. Although this procedure may require
improvements in the determination of the errors, which may be
underestimated, the success
of this fit is apparent from 
Tables~\ref{tab:parameters}, \ref{tab:fitted}.
In turn, this
allows a determination of the CKM matrix with a small
uncertainty in each of the parameters, even smaller than in the general
case discussed in the previous section. This is also shown in
Fig \ref{fig:sins}a, \ref{fig:sins}b, both for the case of free phases
and for the case of maximal phases. As to the value of $m_u/m_d$, this
is essentially unchanged from the general case when the phases are
left free, whereas, for maximal phases, $m_u/m_d=0.606\pm 0.022$.

\section{Conclusions}

We have studied supersymmetric theories of flavor based on a flavor group U(2),
with breaking pattern $\mbox{U(2)} \stackrel{\epsilon}{\rightarrow} 
\mbox{U(1)} \stackrel{\epsilon'}{\rightarrow} 0$, and symmetry breaking
parameters $\epsilon \approx m_s/m_b$ and $\epsilon' \approx \sqrt{m_d
m_s/m_b^2}$. These parameters are sufficiently
small that the quark and lepton mass matrices are dominated by terms up to
linear order in $\epsilon$ and $\epsilon'$, and must therefore arise from just 
4 types of interactions: $\psi_3 \psi_3 H + (1/M)(\psi_3 \phi^a
\psi_a + \psi_a S^{ab} \psi_b + \psi_a A^{ab} \psi_b)H$,
with $S^{ba} = +S^{ab}$ and $A^{ba} = -A^{ab}$. 
Allowing the most general breaking of U(2) to U(1), by $\vev{S^{22}},
\vev{\phi^2}$ of order $\epsilon$, and assuming that the final U(1) is broken
only by $\vev{A^{12}}$ of order $\epsilon'$, a simple symmetry origin is found
for a highly successful texture.
This symmetry structure also solves the  supersymmetric
flavor-changing problem, while strongly suggesting that the exchange of
superpartners at the weak scale will lead to observable rare flavor-changing
and CP violating effects in future experiments. It is
interesting that such a simple symmetry structure simultaneously provides a
very constrained structure for the Yukawa matrices, and an acceptable form for
the scalar mass matrices.

U(2) and its hierarchical breaking,
$\mbox{U(2)}\stackrel{\epsilon}{\rightarrow}  
\mbox{U(1)} \stackrel{\epsilon'}{\rightarrow} 0$, are sufficient to
qualitatively  
understand all the observed small fermion mass ratios and mixing
angles, except  
the large $m_t/m_b$ ratio and the observation that the mass hierarchies in 
the up sector are larger than those in the down and charged lepton sectors.
However, the combination of U(2) and grand unified symmetries allow a symmetry
understanding for the large $m_t/m_c$ and $m_c/m_u$ hierarchies, which
involve a small symmetry breaking parameter, $\rho$, the
ratio of the SU(5) breaking scale to the UV cutoff of the theory. Furthermore,
these symmetries enforce a correlation between these mass hierarchies and the
mass relations $m_\mu = 3 m_s$ and $m_e m_\mu/ m_\tau^2 = m_d m_s/
m_b^2$ at the 
unification scale --  these mass relations are a necessary consequence of
requiring large $m_t/m_c$ and $m_c/m_u$ ratios.
In Tables~\ref{tab:qualitative1} and~\ref{tab:qualitative2} we give
qualitative expressions for the 13 flavor observables 
of the standard model in terms of 4 small parameters $\epsilon, \epsilon',
\rho$ and $\xi$, a Higgs mixing parameter which allows large $m_t/m_b$.
Hence unified U(2) theories give 9 approximate relations. 
Of these 9 relations, 5 are precise, receiving corrections which
are higher order in the symmetry breaking parameters. These are the relations
(15) and (16) for $|V_{ub}/V_{cb}|$ and $|V_{td}/V_{ts}|$, which follow purely 
from the texture dictated by the U(2) symmetry, and the three unified mass 
relations (8), (9) and (10) for $m_b, m_s, m_d$ in terms of $m_\tau, m_\mu, 
m_e$.

These results follow from a general operator analysis and apply in a wide class
of unified U(2) theories. In section~\ref{sec:SU(5)} we described a
class of SU(5) theories, 
while in sections~\ref{sec:direct} and \ref{sec:adjoint}  we discussed
two classes of SO(10) theories, 
distinguished by the SO(10) transformation properties of $\phi^a, S^{ab}$ and
$A^{ab}$. All these theories lead to the same constrained form for the Yukawa
matrices, which successfully accounts for the known quark and lepton masses and
mixings, provided the CKM unitarity triangle is constrained so that $\sin 2
\alpha$ and $\sin 2 \beta$ lie in the shaded (dark+light) region of
Figure~\ref{fig:sins}. 
This is a crucial prediction of the unified U(2) theories.
The unified U(2) theories also predict $m_u/m_d = 0.35$--$0.90$ at 90\%
confidence level.

The light quark and lepton masses, and the non-trivial structure for the CKM
matrix, arise from non-renormalizable operators of the expansion in the
effective theory. In section~\ref{sec:model} we propose a specific
SO(10) model in which 
these non-renormalizable operators are generated by the exchange of a U(2)
doublet of heavy vector generations. This is the simplest unified U(2) theory
that we know. The Yukawa matrices at the unification scale feel SU(5) breaking
only via the mass of the heavy vector generations and the orientations of the
$\phi$ and $S$ vevs, which transform as SO(10) adjoints. When fit to the data,
this theory produces a somewhat tighter prediction for the CKM unitarity
triangle compared to the general unified theories, as seen from the dark 
shaded region of Figure~\ref{fig:sins}.

An interesting feature of this model is that the fit to the data shows that the
three independent physical phases which enter the Yukawa matrices are
constrained to be close to multiples of $\pi/2$, suggesting a spontaneous
origin for CP violation via the vevs which break the flavor and grand unified
symmetries. In this case, the Yukawa matrices depend on just 7
parameters, and a 
fit to data produces precise predictions for $\sin 2 \alpha$, $\sin 2 \beta$
and $|V_{ub}/V_{cb}|$. The scalar mass matrices of this model are more
restricted than in the general unified U(2) theories, allowing a calculation of
the mixing matrices at gaugino vertices. The resulting predictions for
the supersymmetric contributions to flavor and CP violating observables will be
reported elsewhere.

\begin{center}
{\bf Acknowledgments}
\end{center}
This work was supported in part by the Director, Office of
Energy Research, Office of High Energy and Nuclear Physics, Division of
High Energy Physics of the U.S. Department of Energy under Contract
DE-AC03-76SF00098, in part by the National Science Foundation under
grant PHY-95-14797 and in part by the U.S. Department of energy under
contract DOE/ER/01545--700.





\end{document}